\documentclass[twoside,twocolumn,9pt]{article}
\usepackage{extsizes}
\usepackage[super,sort&compress,comma]{natbib} 
\usepackage[version=3]{mhchem}
\usepackage[left=1.5cm, right=1.5cm, top=1.785cm, bottom=2.0cm]{geometry}
\usepackage{balance}
\usepackage{widetext}
\usepackage{times,mathptmx}
\usepackage{sectsty}
\usepackage{graphicx} 
\usepackage{lastpage}
\usepackage[format=plain,justification=raggedright,singlelinecheck=false,font={stretch=1.125,small,sf},labelfont=bf,labelsep=space]{caption}
\usepackage{float}
\usepackage{fancyhdr}
\usepackage{fnpos}
\usepackage[english]{babel}
\usepackage{array}
\usepackage{droidsans}
\usepackage{charter}
\usepackage[T1]{fontenc}
\usepackage[usenames,dvipsnames]{xcolor}
\usepackage{setspace}
\usepackage[compact]{titlesec}

\usepackage{epstopdf}%This line makes .eps figures into .pdf - please comment out if not required.
\usepackage{amssymb}
\usepackage{xr}
\definecolor{cream}{RGB}{222,217,201}

\begin{document}

\pagestyle{fancy}
\thispagestyle{plain}
\fancypagestyle{plain}

%%%%HEADER%%%
%\fancyhead[C]{\includegraphics[width=18.5cm]{head_foot/header_bar}}
%\fancyhead[L]{\hspace{0cm}\vspace{1.5cm}\includegraphics[height=30pt]{head_foot/journal_name}}
%\fancyhead[R]{\hspace{0cm}\vspace{1.7cm}\includegraphics[height=55pt]{head_foot/RSC_LOGO_CMYK}}
%\renewcommand{\headrulewidth}{0pt}
%}
%%%%END OF HEADER%%%

%%%PAGE SETUP - Please do not change any commands within this section%%%
\makeFNbottom
\makeatletter
\renewcommand\LARGE{\@setfontsize\LARGE{15pt}{17}}
\renewcommand\Large{\@setfontsize\Large{12pt}{14}}
\renewcommand\large{\@setfontsize\large{10pt}{12}}
\renewcommand\footnotesize{\@setfontsize\footnotesize{7pt}{10}}
\makeatother

\renewcommand{\thefootnote}{\fnsymbol{footnote}}
\renewcommand\footnoterule{\vspace*{1pt}% 
\color{cream}\hrule width 3.5in height 0.4pt \color{black}\vspace*{5pt}} 
\setcounter{secnumdepth}{5}

\makeatletter 
\renewcommand\@biblabel[1]{#1}            
\renewcommand\@makefntext[1]% 
{\noindent\makebox[0pt][r]{\@thefnmark\,}#1}
\makeatother 
\renewcommand{\figurename}{\small{Fig.}~}
\sectionfont{\sffamily\Large}
\subsectionfont{\normalsize}
\subsubsectionfont{\bf}
\setstretch{1.125} %In particular, please do not alter this line.
\setlength{\skip\footins}{0.8cm}
\setlength{\footnotesep}{0.25cm}
\setlength{\jot}{10pt}
\titlespacing*{\section}{0pt}{4pt}{4pt}
\titlespacing*{\subsection}{0pt}{15pt}{1pt}
%%%END OF PAGE SETUP%%%

%%%%FOOTER%%%
%\fancyfoot{}
%\fancyfoot[LO,RE]{\vspace{-7.1pt}\includegraphics[height=9pt]{head_foot/LF}}
%\fancyfoot[CO]{\vspace{-7.1pt}\hspace{13.2cm}\includegraphics{head_foot/RF}}
%\fancyfoot[CE]{\vspace{-7.2pt}\hspace{-14.2cm}\includegraphics{head_foot/RF}}
%\fancyfoot[RO]{\footnotesize{\sffamily{1--\pageref{LastPage} ~\textbar  \hspace{2pt}\thepage}}}
%\fancyfoot[LE]{\footnotesize{\sffamily{\thepage~\textbar\hspace{3.45cm} 1--\pageref{LastPage}}}}
%\fancyhead{}
%\renewcommand{\headrulewidth}{0pt} 
%\renewcommand{\footrulewidth}{0pt}
%\setlength{\arrayrulewidth}{1pt}
%\setlength{\columnsep}{6.5mm}
%\setlength\bibsep{1pt}
%%%%END OF FOOTER%%%

%%%FIGURE SETUP - please do not change any commands within this section%%%
\makeatletter 
\newlength{\figrulesep} 
\setlength{\figrulesep}{0.5\textfloatsep} 

\newcommand{\topfigrule}{\vspace*{-1pt}% 
\noindent{\color{cream}\rule[-\figrulesep]{\columnwidth}{1.5pt}} }

\newcommand{\botfigrule}{\vspace*{-2pt}% 
\noindent{\color{cream}\rule[\figrulesep]{\columnwidth}{1.5pt}} }

\newcommand{\dblfigrule}{\vspace*{-1pt}% 
\noindent{\color{cream}\rule[-\figrulesep]{\textwidth}{1.5pt}} }

\makeatother
%%%END OF FIGURE SETUP%%%

%%%TITLE, AUTHORS AND ABSTRACT%%%
\twocolumn[
  \begin{@twocolumnfalse}
\vspace{3cm}
\sffamily

\noindent\LARGE{\textbf{Substrate-assisted 2D DNA lattices and algorithmic
      lattices from single-stranded tiles$^\dag$}} \\%Article title goes here
                                %instead of the text "This is the title"
\vspace{0.3cm}  \\

\large{Junghoon Kim,\textit{$^{a}$} Tai Hwan Ha,\textit{$^{b}$}
  and Sung Ha Park\textit{$^{\ast a,c}$}} \\%Author names go here instead of "Full name", etc.

\normalsize{
  We present a simple route to circumvent kinetic traps which affect many types
  of DNA nanostructures in their self-assembly process. Using this method, a new
  2D DNA lattice made up of short, single-stranded tile (SST) motifs was
  created. Previously, the growth of SST DNA assemblies was restricted to 1D
  (tubes and ribbons) or finite-sized 2D (molecular canvases). By utilizing the
  substrate-assisted growth method, sets of SSTs were designed as unit cells to
  self-assemble into periodic and aperiodic 2D lattices which continuously grow
  both along and orthogonal to the helical axis. Notably, large-scale ($\sim$1
  $\mu$m$^2$) fully periodic 2D lattices were fabricated using a minimum of just
  2 strand species. Furthermore, the ability to create 2D lattices from a few
  motifs enables certain rules to be encoded into these SSTs to carry out
  algorithmic self-assembly. A set of these motifs were designed to execute
  simple 1-input 1-output COPY and NOT algorithms, the space-time manifestations
  which were aperiodic 2D algorithmic SST lattices. The methodology presented
  here can be straightforwardly applied to other motifs which fall into this
  type of kinetic trap to create novel DNA crystals.} \\%The abstrast goes here
                                                        %instead of the text
                                                        %"The abstract should
                                                        %be..."

 \end{@twocolumnfalse} \vspace{0.6cm}

  ]
%%%END OF TITLE, AUTHORS AND ABSTRACT%%%

%%%FONT SETUP - please do not change any commands within this section
\renewcommand*\rmdefault{bch}\normalfont\upshape
\rmfamily
\section*{}
\vspace{-1cm}

%%%FOOTNOTES%%%

\footnotetext{\textit{$^{a}$~Department of Physics, Sungkyunkwan University, Suwon 440-746, Korea.}}
\footnotetext{\textit{$^{b}$~Research Center of Integrative Cellulomics, Korea
    Research Institute of Bioscience and Biotechnology (KRIBB), Daejeon 305-806,
    Korea}}
\footnotetext{\textit{$^{c}$~Sungkyunkwan Advanced Institute of Nanotechnology
    (SAINT), Sungkyunkwan University, Suwon 440-746, Korea. E-mail: sunghapark@skku.edu}}

%Please use \dag to cite the ESI in the main text of the article.
%If you article does not have ESI please remove the the \dag symbol from the title and the footnotetext below.
\footnotetext{\dag~Electronic Supplementary Information (ESI) available: [details of any supplementary information available should be included here]. See DOI: 10.1039/b000000x/}
%additional addresses can be cited as above using the lower-case letters, c, d, e... If all authors are from the same address, no letter is required

%\footnotetext{\ddag~Additional footnotes to the title and authors can be included \emph{e.g.}\ `Present address:' or `These authors contributed equally to this work' as above using the symbols: \ddag, \textsection, and \P. Please place the appropriate symbol next to the author's name and include a \texttt{\textbackslash footnotetext} entry in the the correct place in the list.}

%%%END OF FOOTNOTES%%%

%%%MAIN TEXT%%%%
%The main text of the article\cite{Mena2000} should appear here.
% 
%\subsection{This is the subsection heading style}
%Section headings can be typeset with and without numbers.\cite{Abernethy2003}
% 
%\subsubsection{This is the subsubsection style.~~} These headings should end in a full point.  
% 
%\paragraph{This is the next level heading.~~} For this level please use \texttt{\textbackslash paragraph}. These headings should also end in a full point.

\section*{Introduction}

Spatial and temporal control of matter down to the smallest attainable scale is
an ongoing challenge in many fields such as supramolecular chemistry, material
science, and physics. The relatively nascent field of DNA nanotechnology has
provided some proven methods of bottom-up self-assembly to tackle these
problems. One of the core developments in this field has been the creation of
self-assembled DNA crystals in 1-,\cite{liu04, park05, yin08, hamada09}
2-,\cite{winfree98, yan03, ding04, he05, malo05} and 3-\cite{zheng09}dimensions.
The overwhelming majority of these works have utilized the tile-based method
where the oligonucleotides first self-assemble into rigid constituent
building-blocks (tiles) which in turn bind according to their respective
sticky-ends to form crystals during the annealing process. A more direct and
simpler route of creating DNA crystals where extremely simple single strands of
DNA called single-stranded tiles (SSTs), which are effectively made up of only
the sticky-ends thereby bypassing the tile body formation phase, was proposed by
Yin {\em et al}.\cite{yin08} The types of crystals made in that work were those
of the tube and ribbon types, inherently preserving the translational symmetry
to only one direction and restricting the growth of the crystal to 1D, namely
along the helical axis. More recently, SSTs have been programmed to
self-assemble into complex 2 and 3D shapes by way of creating finite-sized
molecular canvases\cite{wei12, ke12} analogous to what DNA
origami\cite{rothemund06, han11} had antecedently achieved.

Despite the successes that have been accomplished in DNA self-assembly, a
chronic problem in the rational design of DNA nanostructures is our lack of
understanding of the self-assembly kinetics. The existence of kinetic traps
which affects the self-assembly kinetics of many structures makes the route
going from structure design to target structure formation a very nontrivial
process.\cite{doye13} In this regard, one prevalent type of kinetic trap which
manifests itself during the self-assembly process is the circularization of
structures to form 1D tubes (such as the aforementioned SST tubes), although
according to their design principles these structures should be able to form
periodic 2D lattices. Examples include motifs such as triple-crossover
tiles,\cite{liu04} cross-tiles,\cite{yan03} DAE-type double-crossover (DX)
tiles,\cite{rothemund04-2, ke06} DX-like structures,\cite{liu06} and
SSTs,\cite{yin08, wei12} all of which show tube formation behavior. Here we
present a method to avoid this type of kinetic trap and use it to create a
significant molecular construct which has been lacking, {\em i.e.}, a periodic
2D SST lattice which retains translational symmetry in both orthogonal
directions allowing full 2D crystal growth. Furthermore, as tile-based
algorithmic self-assembly was an integral advancement in molecular
computing,\cite{mao00, rothemund04, fujibayashi08} we show that these SSTs can
also carry out computations to form 2D algorithmic SST lattices.

\section*{Results and discussion}

\begin{figure}[h]
  \centering
  \includegraphics[scale=0.45]{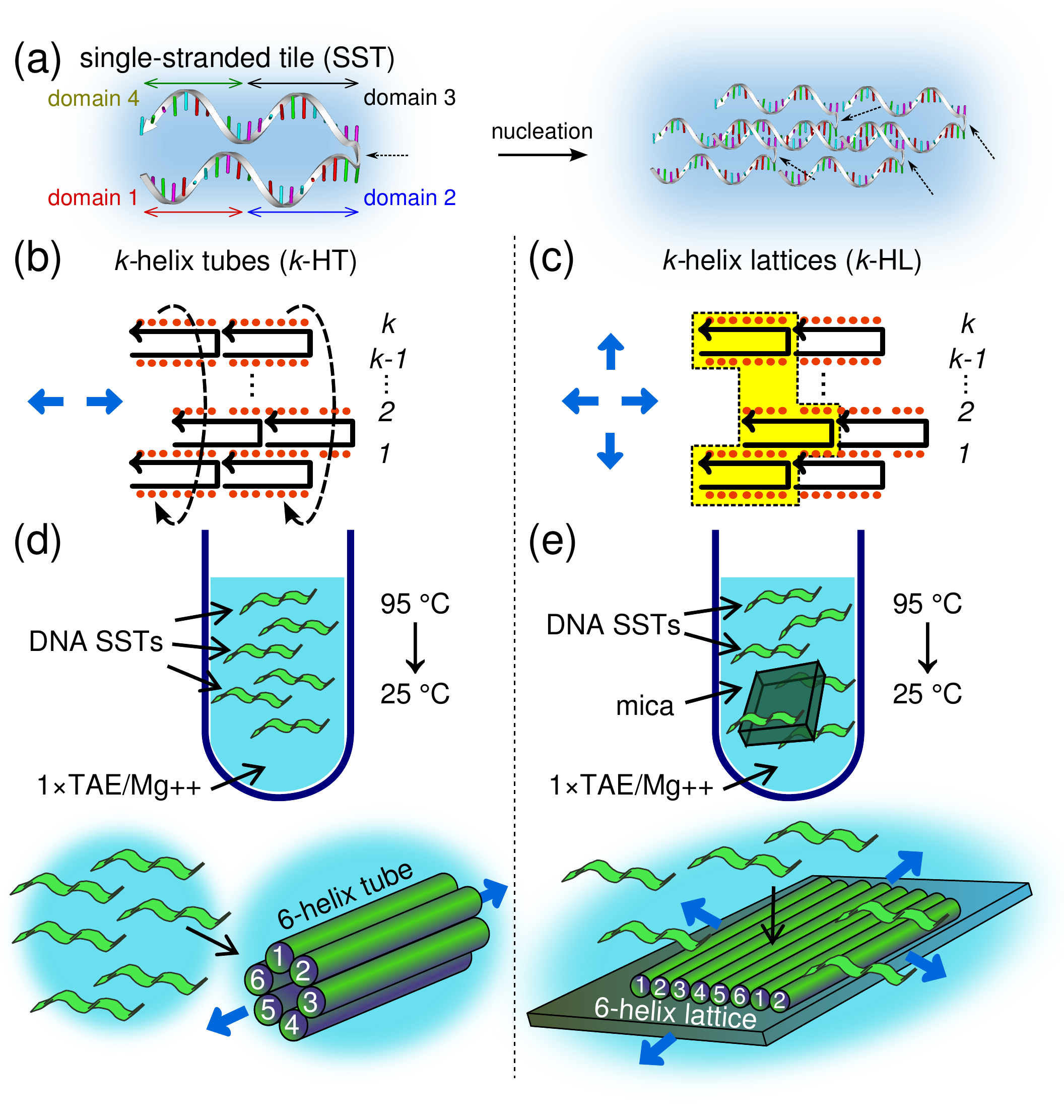}
  \caption{2D structures from SSTs. (a) A single-stranded tile (SST) motif and
    its 4 modular domains. The dotted arrow indicates the half cross-over point.
    Schematics (b and c) and annealing protocols and self-assembly (d and e) of
    $k$-helix tubes ($k$-HTs) and $k$-helix lattices ($k$-HLs), respectively.
    The red dots in (b) and (c) represent base pairings and the black arrow
    heads of the solid lines indicate $3^\prime$ ends. For $k$-HLs, the set of
    $k$ different SSTs delineated by the yellow area in (c) represents a unit
    cell of the 2D lattice and the blue arrows in (b)-(e) represent the crystal
    growth directions.}
  \label{fig1}
\end{figure}

Each SST motif is a 42-base oligonucleotide divided into 4 modular domains. The
modular domains are designed so that between complementary strands, the
odd-numbered domains (domains 1 and 3) bind together and the even-numbered
domains (domains 2 and 4) bind together [Figure~\ref{fig1}a, electronic
  supplementary informtion (ESI) section~\ref{SI:sequencemap}]. Starting from
the 5$^\prime$-end, each strand forms part of a helix up to the first 21 bases
(2 full-turns) and then crosses over where the remaining 21 bases form part of
the adjacent helix in the opposite direction. From this design, we can see these
strands self-assemble into adjoining helices with each adjacent helix having
single-stranded cross-over junctions spaced 21 bases apart. The tubes formed
from this design have circumferences which can be controlled by the number of
different strands, $k$, used. For example, for a tube of $k=6$ (a 6-helix tube
or 6-HT, Figure~\ref{fig1}b), we design 6 different SSTs so that domains 3 and 4
(the top-half) of strand 1 are complementary to domains 1 and 2 (the
bottom-half) of strand 2, respectively, domains 3 and 4 of strand 2 are
complementary to domains 1 and 2 of strand 3, respectively, and so forth up
until the top-half of strand 6 is complementary to the bottom-half of strand 1.
One might (wrongfully) conclude that a set of these $k=6$ strands would act as a
unit cell and lead to 2D lattice formation since, after nucleation, growth is
possible both along and orthogonal to the helical axis. Also, another
theoretically conceivable scenario is the formation of polydisperse $mk$-helix
tubes (where $m\geq2$, $m\in \mathbb{Z}$, {\em e.g.,} 12-helix tubes for $m=2$
and $k=6$). Instead, Yin {\em et al}. postulated that the energy landscape is
such that the tubes are trapped at a free energy local minimum in which
monodisperse 6-HTs are highly favored, terminating any growth orthogonal to the
helical axis once bindings occur to close the tubes.\cite{yin08} For continued
crystal growth orthogonal to the helical axis, this cyclization must somehow be
prevented by offsetting this kinetic trap. Although the origins of this kinetic
trap have not been experimentally verified, one plausible possibility is the
existence of a nucleation barrier between $k$-helix tubes and $mk$-helix tubes
which would naturally explain the monodispersity of the tube circumferences for
a given $k$.

\begin{figure*}[!t]
  \centering
  \includegraphics[scale=0.4]{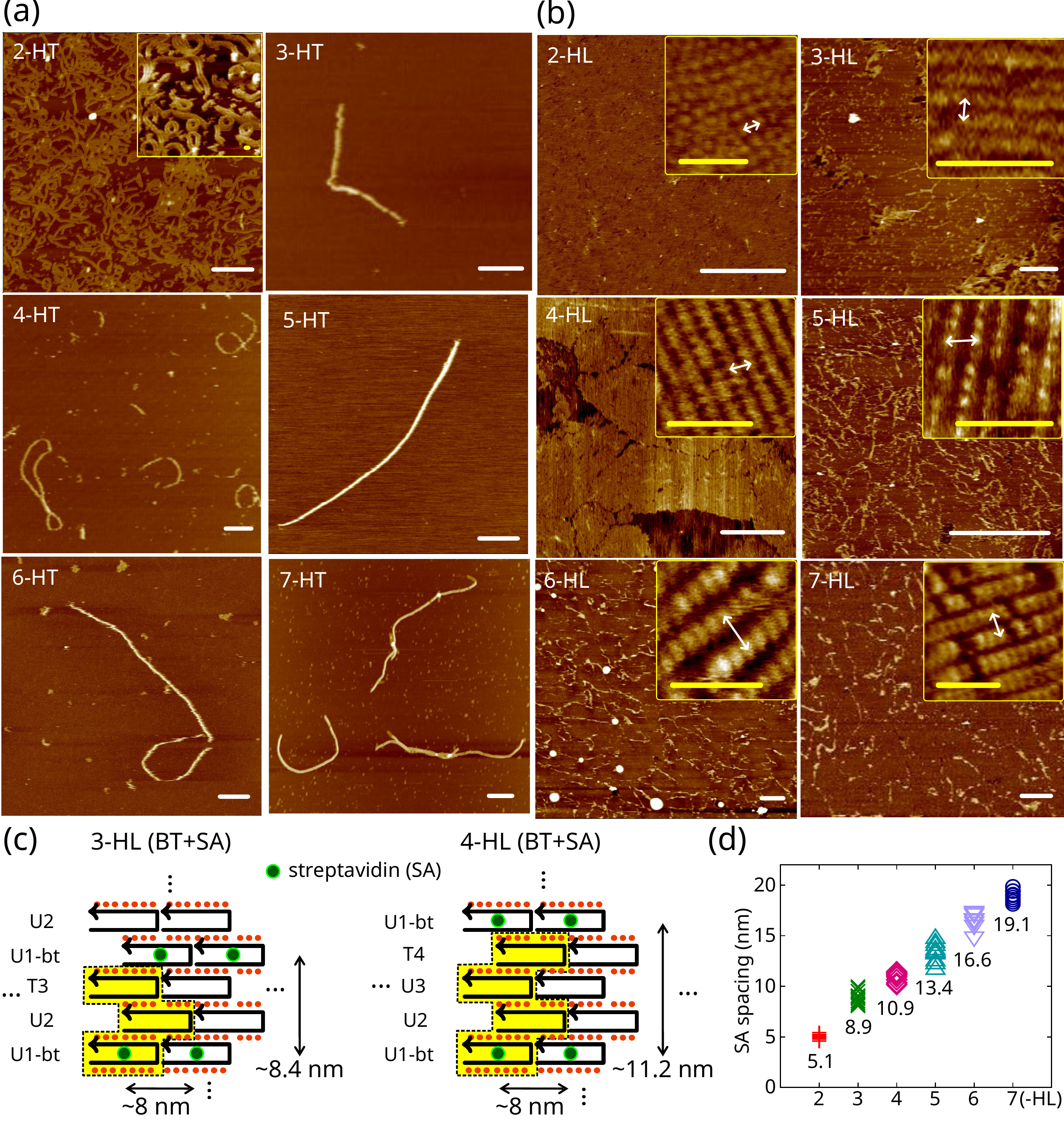}
  \caption{AFM images of the (a) tubes and (b) lattices. Insets show magnified
    views of $k$-HT/HL structures and the white arrows indicate typical spacings
    between neighboring streptavidin (SA) sites orthogonal to the helical axis,
    {\em i.e.}, the width of a unit cell for each $k$-HL (see ESI
    section~\ref{SI:sequencemap} for information on biotinylated (BT) sites and
    ESI Figure~\ref{SIfig:AFM2} for distance measurements between streptavidin
    molecules). The periodicity of the lattice can be clearly seen in all the
    helix lattices. (c) Two examples, 3-HL \& 4-HL, are shown with their
    respective unit cells outlined in a dashed yellow area. Arrows denote
    spacings between SA sites. (d) The average of 10 measured spacings between
    nearest-neighbor SA sites orthogonal to the helical axis like the ones
    indicated by the white arrows in (b), for each $k$-HL. (White scale bars :
    500 nm; Yellow scale bars : 50 nm)}
  \label{fig2}
\end{figure*}

Our strategy is to remove this barrier so growth can continue and form 2D
lattices which preserve translational symmetry in both directions
(Figure~\ref{fig1}c). This can be achieved by the substrate-assisted growth
(SAG) method, where a small sheet of mica (or other substrates with siloxy
groups at its surface such as fused silica or quartz) is inserted into the
solution vessel at the start of the annealing process
(Figure~\ref{fig1}e).\cite{hamada09} The addition of the mica substrate into the
solution provides preferential nucleation sites which significantly reduces the
kinetic barrier to nucleation (as compared to free-solution annealing) and
changes the nucleation mechanism from a homogeneous to a heterogeneous one. This
is consistent with observations from previous studies of SAG methods where the
saturation concentration, {\em i.e.}, the concentration at which nucleated seeds
start to form, was found to be reduced by roughly an order of
magnitude.\cite{lee13} By offsetting the kinetic trap, adjoining helices are no
longer forced into tubular structures (Figure~\ref{fig1}d), {\em i.e.},
$k$-helix tubes ($k$-HT), but rather form well-defined 2D lattices, {\em i.e.},
$k$-helix lattices ($k$-HL), on the mica surface (Figure~\ref{fig1}e).
Figure~\ref{fig2} shows atomic force microscopy (AFM) data of the structures
assembled from SSTs via free-solution annealing and substrate-assisted annealing
for $k=2$ through 7, respectively. As can be seen from the images, 2D lattices
possessing translational symmetry in both orthogonal directions were
successfully fabricated going as far as using just 2 strand species ($k=2$).
Although the kinetics of the $k=2$ system favors the formation of 2-helix wide
linear chains (2-HT) under free-solution annealing, evidence strongly suggests
that when mica is added to the system pre-annealing, the assembly pathway is
altered such that adjoining helical lattices are highly favored (see ESI
Figure~\ref{SIfig:AFM} and ESI Figure~\ref{SIfig:AFM2} for AFM measurements). It
is worth noting that in a work by Liu {\em et al.},\cite{liu06} 1 strand specie
was used to form DX-like tiles, but assembly of these tiles also fall into the
trap of rolling up to form tubes instead of 2D lattices. The authors of that
work assert that the tubes are kinetically favored over 2D lattices thus keeping
the translational symmetry in only one direction. To check 2D lattice formation
in our work, the first strand (U1-bt) of each $k$-HL set was biotinylated at a
specified site (Figure~\ref{fig2}c). If 2D lattices do indeed form for all $k$,
the spacings between these biotinylated (BT) sites running orthogonal to the
helical axis would increase proportional to $k$ and could be measured by AFM
when bound to streptavidin (SA). Figure~\ref{fig2}b clearly shows 2D periodic
arrangements of SA sites with increasing spacings proportional to $k$ for all
$k$. Figure~\ref{fig2}d shows measurements of 10 randomly chosen pairs of
neighboring SA sites orthogonal to the helical axis taken from AFM images for
each $k$ along with their averages (see also ESI Figure~\ref{SIfig:AFM2}). The
data is in good agreement with predictions given that a single DNA duplex has a
measured width of $2.5\sim 2.8$ nm (ESI Figure~\ref{SIfig:AFM}), {\em e.g.}, the
predicted SA spacing for 4-HL is $4\times 2.6~\text{nm}=10.4~\text{nm}$ whereas
the measured average is 10.9 nm. For all values of $k$ tested, the surface
morphology of $k$-HLs remained the same at 50 nM strand concentrations, {\em
  i.e.}, fully covered single-layered 2D helical lattices on a mica substrate
(see ESI Figure~\ref{SIfig:frac} for an analysis of the fractional coverage
dependence on strand concentration).

Another important aspect we addressed in this work was the creation of
algorithmic 2D crystals from SSTs. We will compare SSTs with one of the most
popular motifs used for algorithmic self-assembly, the DX
tile,\cite{fu93,winfree98,rothemund04} since the geometric topology of both
motifs can be abstracted in the same manner, {\em i.e.}, both have 4 sticky-ends
acting as Wang tile edges.\cite{wang61,wang62} In principle, there are several
advantages of implementing algorithms using SST motifs. First, the information
density of an SST crystal is more than three times higher than DX crystals. The
theoretical area of a single DX motif is $\sim$50.32 ($= 12.58\times 4$) nm$^2$
whereas for an SST motif it is $\sim$14.28 ($= 7.14\times 2$) nm$^2$. This
provides a more intricate platform with a higher resolution in which algorithms
can be carried out compared to tile-based algorithmic self-assembly. Second,
within the kinetic Tile Assembly Model,\cite{winfree98a} the longer sticky-ends
of SSTs allows for much more favorable ``correct'' associations between motifs,
thereby reducing the overall error rate of the algorithmic crystal. More
specifically, at equilibrium, a correct tile association is favored over an
incorrect tile association (an error) by a factor of $e^{G_{\text{se}}}$, where
$G_{\text{se}}$ is the amount of free energy needed to break a single sticky-end
bond and is directly proportional to the sticky-end length. Hence, with all
other factors being equal, correct sticky-end bindings between SST motifs, which
have two different lengths of 10 and 11 nucleotides, are exponentially more
likely compared to sticky-end bindings between the 5-nucleotide sticky-ends of
DX tiles ($e^{G_{\text{se}}^{\text{SST}}} / e^{G_{\text{se}}^{\text{DX}}}$,
where $e^{G_{\text{se}}^{\text{SST}}}$ and $e^{G_{\text{se}}^{\text{DX}}}$ are
the ratios of correct tile associations over error associations of the SSTs and
DX tile sticky-ends, respectively, and are directly proportional to their
respective sticky-end lengths). Third, the longer sticky-ends also allow a much
bigger design space (by more than 3 orders of magnitude), thereby avoiding
unwanted bindings among similar sticky-ends. Also, to a lesser extent, by
circumventing the body formation phase, spurious bindings which may occur during
body formations and their erroneous crystal offshoots which follow, can be
avoided.

\begin{figure}[!h]
  \centering
  \includegraphics[scale=0.34]{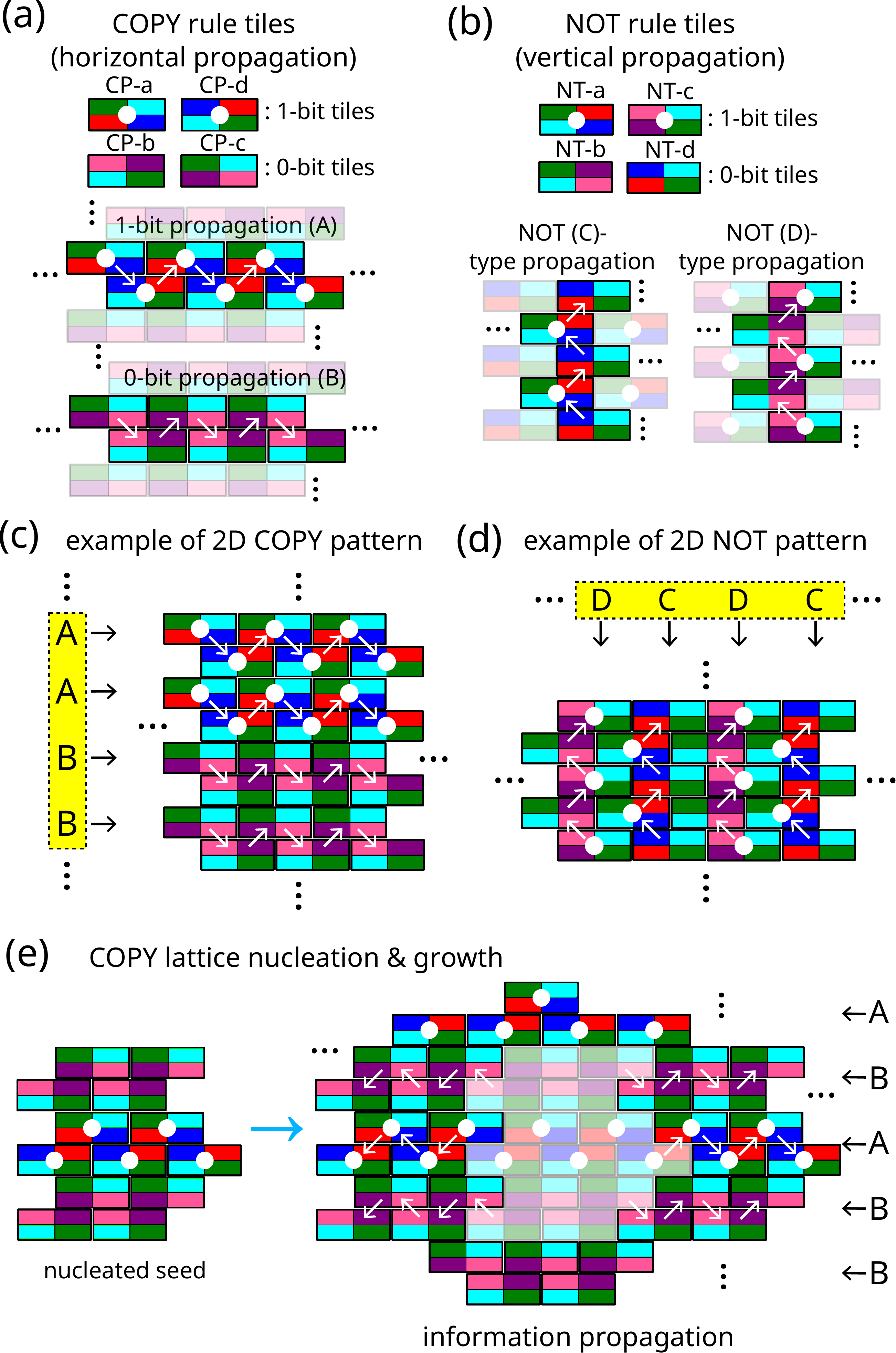}
  \caption{Algorithmic self-assembly of 2D SST lattices. Abstractions of SST
    motifs as 4-color rectangles (a-e) with each domain color-coded where
    matching top and bottom edge colors represent complementarity. Each
    rectangle represents either a 0-bit or 1-bit tile. Information is propagated
    along and perpendicular to the helical axis for (a) COPY and (b) NOT
    patterns, respectively. There are 2 common domains (green and cyan) in every
    tile that do not carry any information but allow 2D lattice growth. 4 SST
    rule motifs used for a 2D (a) COPY and (b) NOT lattices. For COPY patterns,
    the two 1-bit SSTs ({\em i.e.}, CP-a and CP-d) and the two 0-bit SSTs ({\em
      i.e.}, CP-b and CP-c), hybridize according to their respective sticky-ends
    to copy information and produce ``A''- and ``B''-type patterns,
    respectively. For NOT patterns, information is propagated perpendicular to
    the helix axis such that 0-bit and 1-bit tiles form an alternating pattern.
    (c) A 2D COPY crystal can be thought of as randomly repeating A and B
    patterns, such as the $\cdots$AABB$\cdots$ pattern shown here. Each row of
    helices consisting of information encoded sticky-ends are separated by a row
    of helices consisting of common domains. (d) An example of a 2D NOT pattern.
    The direction of information propagation can be in the opposite direction to
    the one specified here for (a-d). (e) Nucleation and growth of a 2D COPY
    lattice. Information propagates away from the seed.}
  \label{fig3}      
\end{figure}

Here we show two primitive 1-input 1-output logic operations, COPY and NOT, in
the form of SST motifs (Figure~\ref{fig3}). Figure~\ref{fig3}a and b show
individual SST motifs used for each logic operation abstracted as 4-color
rectangles, where each domain of each motif is color-coded. Matching colors of
the top and bottom edges of the rectangle represent complementarity and bindings
occur through these edges. For both COPY and NOT operations, 4 SSTs, 2 0-bit and
2 1-bit tiles, were designed to carry out each algorithm. Of the 4 sticky-ends
of each SST, 2 sticky-ends were encoded as inputs/outputs and 2 were designed to
be common in all 4 of the motifs. These common sticky-ends (cyan and green
colored domains) are meant not to propagate information but to act as binding
domains so that the algorithmic lattice can grow beyond the 1 dimension which is
sufficient for 1-input 1-output type algorithmic crystals
(Figure~\ref{fig3}a-e). Since both COPY and NOT operations are reversible, each
information carrying sticky-end can act as both an input and output. The 4 COPY
and NOT SSTs are labeled as CP-$x$ and NT-$x$, respectively, where $x=$a, b, c,
or d. CP-a, -d (NT-a, -c) represent 1-bit tiles and CP-b, -c (NT-b, -d)
represent 0-bit tiles for COPY (NOT) logic gates. To experimentally
differentiate 1-bit and 0-bit SSTs, 1-bit strands were biotinylated at the
appropriate locations (ESI, section~\ref{SI:sequencemap}) so that they could
bind with SA molecules which were added just before AFM imaging.

Once nucleation occurs, crystal growth is dictated by the Tile Assembly Model.
Much akin to algorithmic self-assembly of DX tiles, algorithmic lattice growth
happens when 2-bindings of the SSTs are thermodynamically much more favored than
1-bindings. In addition, since SSTs have much longer sticky-ends than DX tiles,
error rates can be expected to be substantially lower than DX algorithmic
crystals. For COPY lattices, we label the propagation of 1-bit information as
``A'' and 0-bit information as ``B'' (Figure~\ref{fig3}a). Analogously, for NOT
crystals, where an input of a 0-bit motif leads to a binding with a 1-bit motif
(and {\em vice-versa}), there are 2 types of information propagating bindings,
types ``C'' and ``D'' (Figure~\ref{fig3}b). Any random combination of A's and
B's or C's and D's constitute a single COPY or NOT 2D algorithmic lattice,
respectively, of which one example of each pattern is shown in
Figure~\ref{fig3}c and d. On a side note, the direction of information
propagation can only be determined insofar as the crystal growth direction can
be determined and cannot be further specified so the information propagation
(white) arrows in Figure~\ref{fig3}(a-d) may just as well point in the opposite
direction. Figure~\ref{fig3}e shows the direction of information propagation
after nucleation for a COPY lattice. Once a nucleated seed forms, the direction
of information propagation points away from the seed, {\em i.e.}, parallel
(orthogonal) to the helical axis for the COPY (NOT) tiles designed in this work.

\begin{figure}[!h]
  \centering
  \includegraphics[scale=0.5]{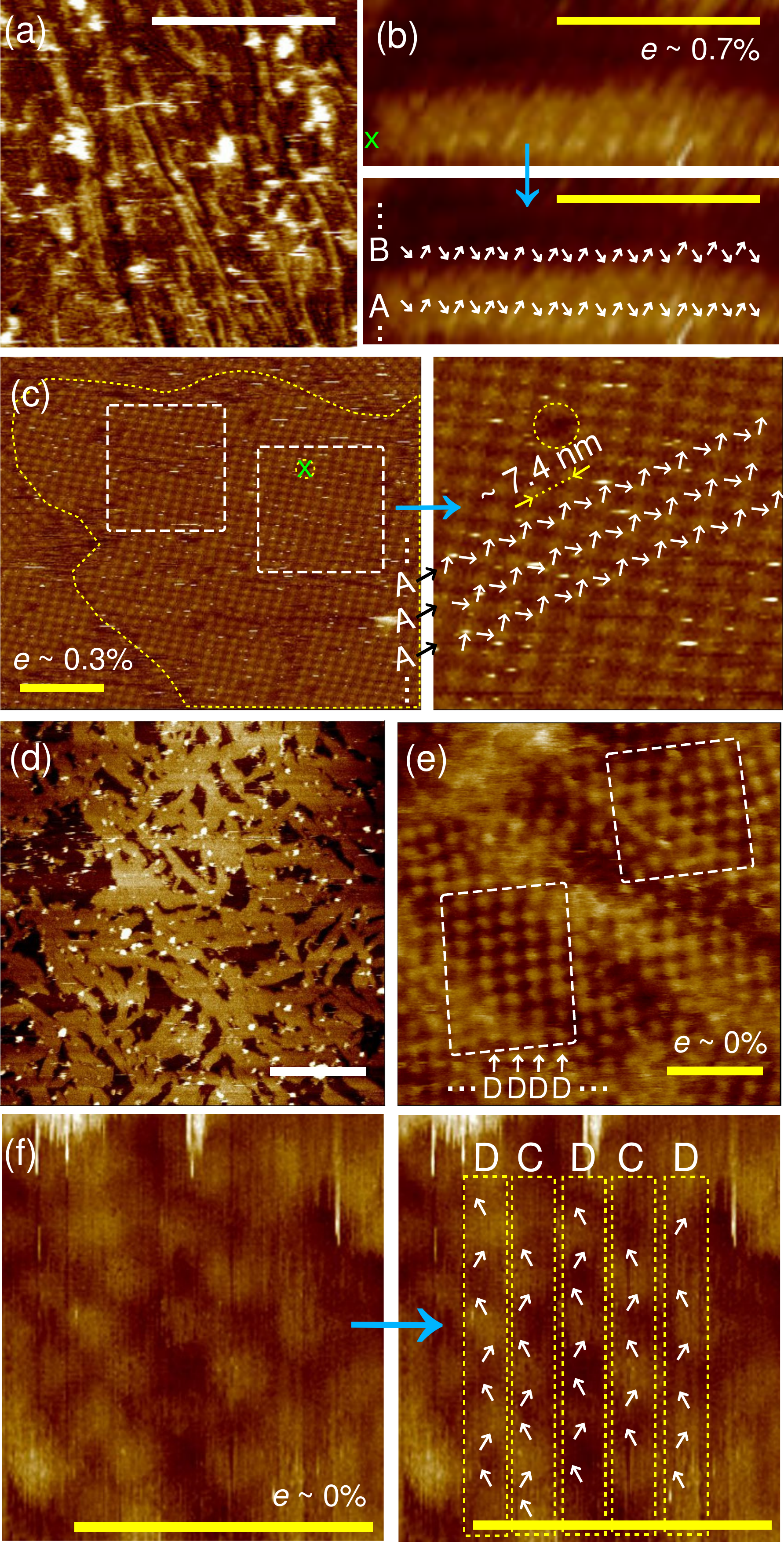}
  \caption{AFM images of SST algorithmic self-assembly. (a) AFM image of a COPY
    lattice. (b) Magnified region showing A- and B-type propagation. (c) A
    biased COPY lattice consisting of A-type propagations. The dashed yellow
    circle indicates either an error or an unbound streptavidin site (probably
    the latter since all the neighboring sites do not deviate from the intended
    pattern). (d) AFM image of a 2D NOT lattice. (e) A biased NOT lattice
    consisting of all D-type (or all C-type, the distinction is not possible)
    propagation. (f) A portion showing a $\cdots$DCDCD$\cdots$ pattern as
    illustrated in Figure~\ref{fig3}d. Errors (green x's) and error rates, $e$,
    for some of the images [whole image : (b) and (f), image sections delineated
      by dashed white lines : (c) and (e)] are shown. (White scale bars : 500
    nm; Yellow scale bars : 50 nm)}
  \label{fig4}  
\end{figure}

Typically observed crystal sizes were on the scale of several hundred nm$^2$ to
$\sim 1$ $\mu$m$^2$ for both periodic and aperiodic crystals, but the crystal
sizes were likely limited by the SAG method since the lack of control of the
nucleation sites on the substrate surface restricts a single crystal from
growing beyond a certain size before encountering other nucleation seeds or
crystals on the surface [Figure~\ref{fig4}a (COPY) and d (NOT), with detailed
  analysis of magnified regions shown in Figure~\ref{fig2}b (COPY) and f (NOT)].
Hence, it is sometimes difficult to verify the exact boundaries of a single
crystal on a fully covered mica surface as partially formed crystals seem to
join together at some of the lattice edges. Some of the formed lattices showed
biased formations where whole lattices were made up of solely A- or B-type (C-
or D-type) propagations for COPY (NOT) lattices (Figure~\ref{fig2}c and e),
which is a possibility due to the design of the tiles. This suggests that there
may be a difference in binding affinities of the tiles making up A- and B-type
(C- and D-type) propagations. Large portions of these biased lattices showed
very few, if any, errors. Figure~\ref{fig2}c (right) shows a magnified region of
a biased COPY lattice (delineated by the dashed white lines) where a very
periodic arrangement of SA sites with spacings of $\sim7.4$ nm can be observed
(A-type propagations). One site, indicated by the dashed yellow circle, may
indicate an error binding but it is more likely that a SA molecule did not bind
at that site or was displaced by the AFM tip during imaging.

\section*{Conclusions}

We have shown here how to avoid a prevalent type of kinetic trap in DNA
self-assembly and were successful in creating novel periodic 2D SST lattices
hundreds of nm$^2$ to $\sim 1$ $\mu$m$^2$ from a small set of $k$ strands (with
a minimum of $k=2$) using this method. Motivational differences aside, creating
2D DNA canvases of similar sizes requires upwards of several hundred SST strand
species (up to 777 for a 2D rectangular canvas) making large-scale 2D lattices
rather cumbersome and inefficient.\cite{wei12} Moreover, the SAG method allows
for algorithmic self-assembly using the simplest of DNA motifs, a single
oligonucleotide, with the possibility to create highly complex, fully
addressable patterns. Although some preliminary theoretical and experimental
studies of SAG DNA nanostructures exist,\cite{hamada09,sun09,lee11,hamada12}
elucidating the precise assembly pathway remains a vital next step in further
exploiting any surface-assisted DNA self-assembly. This is important if we
consider the vast potential DNA structures may hold in modern electronic
applications, a sizable portion which is based on technologically important
materials with surface siloxy groups, {\em e.g.}, silica.\cite{lee11} From our
study, it seems that under conventional annealing protocols in a
$1\times$~TAE/Mg$^{2+}$ buffer, non-specific adsorption of the SSTs onto the
mica surface after which 2D diffusion-based self-assembly occurs is favored over
3D diffusion-based (free-solution) self-assembly and adsorption. By providing
favorable nucleation sites, the mica substrate traps the free floating
oligonucleotides near the surface (within the Debye length) via the Mg$^{2+}$
counter-ions and expedites the reactions of self-assembly of these trapped DNA
strands by reducing the diffusion dimensionality from 3D to 2D. The AFM data
corroborates that the kinetics of these reactions are faster than the two-step
reaction of 3D diffusion-based self-assembly and adsorption onto the mica
surface. This type of reduction in the degree of freedom (diffusion
dimensionality) has been cited as the cause of accelerations in reaction rates
of receptor-ligand interactions.\cite{adam68} Experimentally verifying the
assembly pathway may be possible by obtaining the surface thermal profiles of
SAG SST assemblies in combination with real-time fractional coverage
measurements during the annealing process. Lastly, the method presented here can
be straightforwardly applied to other tile-body-based DNA motifs which
circularize to form tube/ribbon structures during free-solution
annealing\cite{yan03,liu04,rothemund04-2,ke06} to elucidate whether the origins
of the circularization are due to the same kinetic trap as the SSTs studied here
with the further potential to create novel DNA structures.

\section*{Experimental}

\subsection*{DNA oligonucleotide synthesis.}
Synthetic oligonucleotides were purchased from Bioneer Co. Ltd (Daejeon, Korea)
and purified by high performance liquid chromatography (HPLC). Details can be
found at \texttt{www.bioneer.com}.

\subsection*{Annealing protocol.}
Stoichiometric quantities of each strand species of the $k$-HTs and $k$-HLs were
pipetted into AXYGEN-tubes along with a physiological buffer,
1$\times$~TAE/Mg$^{2+}$ [Tris-Acetate-EDTA (40 mM Tris, 1 mM EDTA (pH 8.0), 12.5
  mM Mg(Ac)$_2$)]. The microtubes were then shaken for 30 seconds using a vortex
mixer and centrifuged at 8,000 rpm for 10 seconds. For the substrate-assisted
growth method of the $k$-HLs/COPY/NOT samples, a piece of $5\times 5$~mm$^2$
mica sheet (Pelco\textregistered~mica sheets, Ted Pella, Inc.) was placed inside
the microtube after centrifuging. The samples were then cooled slowly from 95
$^{\circ}$C to 25 $^{\circ}$C by placing the AXYGEN-tubes in 1.5 L of boiled
water in a styrofoam box for 24 hours to facilitate hybridization. The strand
species concentrations were 200 nM and 50 nM and the final volumes were 100
$\mu$l and 200 $\mu$l for the $k$-HTs and $k$-HLs/COPY/NOT samples,
respectively.

\subsection*{AFM imaging.}
The AFM images of the $k$-HTs were taken by pipetting 5 $\mu$L of the samples on
freshly cleaved mica after which 45 $\mu$L of 1$\times$~TAE/Mg$^{2+}$ buffer was
pipetted onto the mica surface and another 5 $\mu$L of 1$\times$~TAE/Mg$^{2+}$
buffer was dropped onto the AFM tip (Veeco Inc.). For the $k$-HLs/COPY/NOT
samples, the mica sheets were taken out the microtubes and one side blow dried
with nitrogen gas after which superglue was lightly applied on the dry side and
placed onto a metal puck (Ted Pella, Inc.). 20 $\mu$L of 1$\times$~TAE/Mg$^{2+}$
buffer was pipetted onto the mica and the metal puck placed on the AFM scanner
head for imaging. For biotinylated $k$-HLs/COPY/NOT samples, we added 5 $\mu$L
of 200 nM streptavidin (Rockland Inc.) to the mica surface and let the sample
sit for 1 minute before imaging. All AFM images were obtained on a Digital
Instruments Nanoscope III (Vecco, USA) with a multimode fluid cell head in
tapping mode under a buffer using NP-S oxide-sharpened silicon nitride tips
(Vecco, USA).

\section*{Acknowledgements}
This research was supported by a grant from the Korea Research Institute of
Bioscience and Biotechnology (KRIBB) Research Initiative Program (2014) and by
the National Research Foundation of Korea (NFR) funded by the Ministry of
Science, ICT \& Future Planning (MSIP) (NRF-2014R1A2A1A11053213).

\nocite{*}

%%%REFERENCES%%%
\bibliography{SST} %You need to replace "rsc" on this line with the name of your .bib file
\bibliographystyle{rsc} %the RSC's .bst file

%%%%%%%%%%%%%%%%%%%%%%%%%%%%%%%%%%%%%%%%%%%%%%%%%%%%%%%%%%%%%%%%%%%%%%%%%%%%%%%%
%% TOC
%%%%%%%%%%%%%%%%%%%%%%%%%%%%%%%%%%%%%%%%%%%%%%%%%%%%%%%%%%%%%%%%%%%%%%%%%%%%%%%%
\clearpage

\setcounter{figure}{0}
\renewcommand{\figurename}{TOC}
\renewcommand{\thefigure}{}

\begin{figure*}
 \centering
  \includegraphics[scale=0.5]{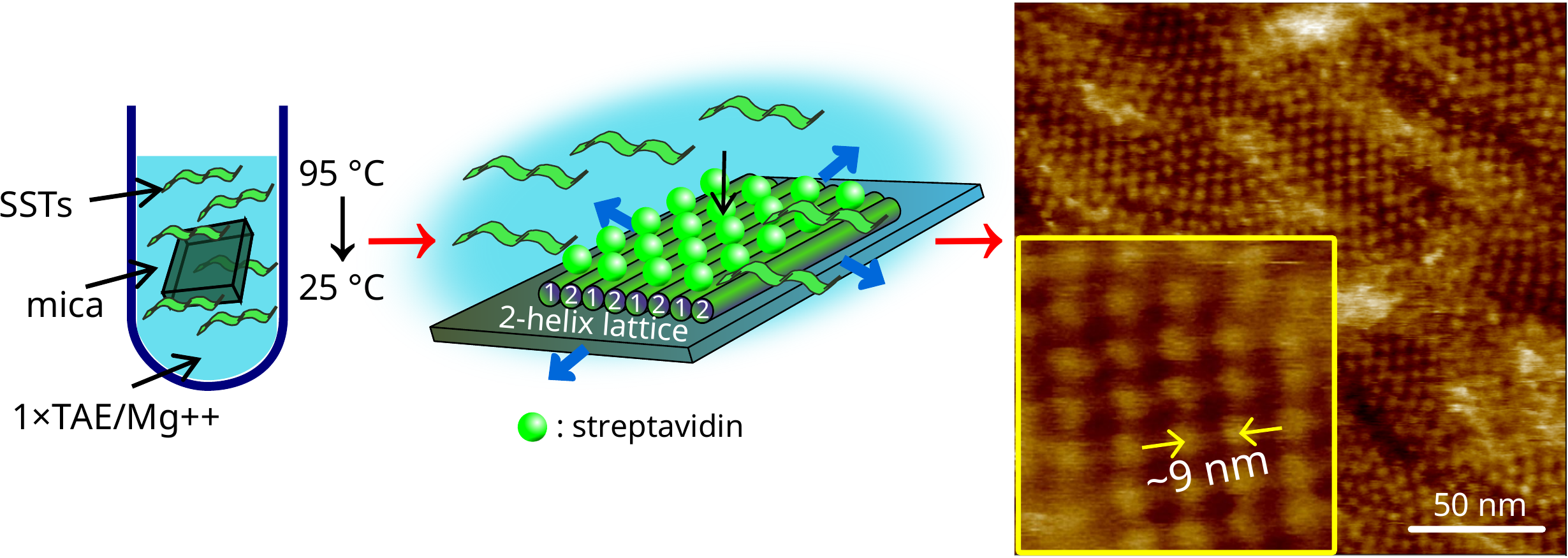}
  \caption{We demonstrated large-scale periodic and algorithmic 2D DNA crystals
    created from single-stranded tiles.\\ ({\bf Keywords :} single-stranded
    tile, DNA structures, substrate-assisted growth, self-assembly, algorithmic
    self-assembly)}
  \label{TOC}    
\end{figure*}

\end{document}